\documentclass[11pt]{article}
\usepackage[a4paper,lmargin=2.5cm,rmargin=2.5cm,tmargin=4cm,bmargin=3.5cm]{geometry}
\usepackage[utf8x]{inputenc}
\usepackage{dsfont}
\usepackage{amsmath,amssymb,amstext,amsthm,amscd, amsxtra, mathrsfs,bm,slashed}
\usepackage{nicefrac}
\usepackage{enumitem}
\usepackage{array}
\usepackage{cite}
\usepackage{todonotes}
\usepackage{mathtools}
\usepackage[affil-it]{authblk}

 \usepackage{hyperref}
 \hypersetup{
   pdftitle   = {LTE states in QFT},
  pdfauthor  = {M. Gransee},
 }

\def\ben{\begin{equation}}
\def\een{\end{equation}}

\def\cA{{\mathcal A}}

\def\cC{{\mathcal C}}
\def\cD{{\mathcal D}}

\def\cF{{\mathcal F}}

\def\cO{{\mathcal O}}

\def\cQ{{\mathcal Q}}

\def\cS{{\mathcal S}}
\def\cT{{\mathcal T}}

\def\IC{{\mathbb C}}
\def\IM{{\mathbb M}}
\def\IN{{\mathbb N}}
\def\IR{{\mathbb R}}
\def\IZ{{\mathbb Z}}

\def\a{\alpha}
\def\b{\beta}
\def\g{\gamma}        
\def\d{\delta}        
\def\eps{\varepsilon}

    \def\Th{\Theta}

       \def\L{\Lambda}
\def\m{\mu}

\def\x{\xi}

\def\s{\sigma}		 
    
\def\t{\tau}

\def\o{\omega}        

\def\fS{{\mathfrak S}}

		\def\fu{{\mathfrak{u}}}

		\def\fw{{\mathfrak{w}}}

\def\to{\rightarrow}

\def\del{\partial}

\newcommand{\abs}[1]{{\left|{#1}\right|}}			
\newcommand{\eins}{{\mathds{1}}}				
\newcommand{\WF}{{\mathrm{WF}}}				
\newcommand{\wfou}{{\hat{\fw}}}				

\newcommand{\diff}{{\mathrm{d}}}				

\def\bb{{\boldsymbol{\b}}}					

 \newtheorem{thm}{Theorem}[section]

 \newtheorem{prop}[thm]{Proposition}
\newtheorem*{obs}{Observation}
 \theoremstyle{definition}
 \newtheorem{defn}[thm]{Definition}
 \theoremstyle{remark}

\numberwithin{equation}{section}

\begin{document}

\title{Local Thermal Equilibrium States\\ in Relativistic Quantum Field Theory}
\author[1,2]{Michael Gransee}
\affil[1]{MPI f\"ur Mathematik in den Naturwissenschaften, 04103 Leipzig, Germany}
\affil[2]{Institut f\"ur Theoretische Physik, Universit\"at Leipzig, 04103 Leipzig, Germany}
\date{}
\maketitle

\begin{abstract}
It is well-known that thermal equilibrium states in quantum statistical mechanics and quantum field theory can be described in a mathematically rigorous manner by means of the so-called Kubo-Martin-Schwinger (KMS) condition, which is based on certain analyticity and periodicity properties of correlation functions. On the other hand, the characterization of non-equilibrium states which only locally have thermal properties still constitutes a challenge in quantum field theory. We discuss a recent proposal for characterization of such states by a generalized KMS condition. The connection of this proposal to a proposal by D. Buchholz, I. Ojima and H.-J. Roos for characterizing local thermal equilibrium states in quantum field theory is discussed
\end{abstract}

\maketitle
\section{Introduction}
Soon after the introduction of the algebraic approach to quantum field theory, developed by Araki, Haag and Kastler in the 60's \cite{Ara09, Haa92}, it became clear that this framework allows for an immediate adoption to non-relativistic quantum systems, for example spin lattice models. This led to the conclusion that equilibrium states in quantum statistical mechanics should be described in the operator-algebraic framework by the Kubo-Martin-Schwinger (KMS) condition, which was first envisaged by Haag, Hugenholtz and Winnink in \cite{HHW67}. The mathematically rigorous formulation of equilibrium quantum statistical mechanics based on the KMS condition has offered many insights into the structural properties of equilibrium states and at the same time has revealed previously unexpected connections to pure mathematics, in particular to Tomita-Takesaki modular theory, which had a huge influence on the understanding of quantum field theory, see e.g. the review article of Borchers \cite{Bor00}. For references and an extensive discussion of non-relativistic quantum statistical mechanics in the operator algebraic formulation, the reader is referred to the  monograph by Bratteli and Robinson \cite{BR97}.  Interestingly, it took almost 20 years until the KMS condition was used in the rigorous investigation of thermal properties of relativistic quantum fields. This was initiated by Buchholz and Junglas \cite{BJ85, BJ89}, leading to a fully relativistic version of the KMS condition \cite{BB94} and an axiomatic approach to thermal field theory (\`a la Wightman \cite{SW00}), in which the relativistic spectrum condition is replaced by the relativistic KMS condition \cite{BB96}.

Although the KMS condition turned out to be fruitful in this respect, it is clear from the outset that, in general, an arbitrary state of a quantum system will not be an equilibrium (KMS) state, since in nature there also arises a variety of non-equilibrium states ranging from mild perturbations of equilibrium states to steady states (e.g. a steady heat flow through a metal bar) and hydrodynamic flows (for example water in a pipe), up to states which do not admit any thermal interpretation at all. On the side of relativistic QFT, Buchholz, Ojima and Roos \cite{BOR02} developed a method for distinguishing between states which are out of equilibrium but locally still have a thermodynamical interpretation. Heuristically speaking, a \emph{local thermal equilibrium (LTE) state} is defined as a state for which certain (point-like) observables, representing intensive thermal quantities like temperature, pressure and thermal stress-energy, take the same values as they take if the quantum field is in some thermal reference state (a KMS state or a mixture of such).  Below we will discuss how this can be made precise and review several aspects of the LTE condition in quantum field theory.

The KMS condition is based on given analyticity and periodicity properties of correlation functions and yields an \emph{intrinsic} characterization of equilibrium states. In contrast, the LTE condition of Buchholz, Ojima and Roos has to be regarded an \emph{extrinsic} condition, since it is based on the comparison of a state with the members of an \emph{a priori} fixed family of thermal reference states. It seems to be natural to ask if one could characterize such local equilibrium states in a manner similar to the KMS condition, i.e. by an \emph{intrinsic} condition also based on analyticity and periodicity properties of the correlation functions of LTE states. In fact, recent results of Gransee, Pinamonti and Verch show that this is possible. Motivated by the analysis of correlation functions of KMS states of the free quantized scalar field, in \cite{GPV15} a generalized version of the KMS condition, called \emph{local KMS} (LKMS) condition, is introduced. Following this, it is shown that a certain class of LTE states in the sense of \cite{BOR02} can be equivalently described by this condition. We will discuss the LKMS condition and its relation to the LTE condition in section 3 below.

\subsection{Preliminaries}
\paragraph{The QFT model}
For simplicity we consider an uncharged free scalar quantum field on Minkowski spacetime $\IM=\IR^4$, with the Minkowski pseudo-metric $\eta$ of diagonal form $\eta=\text{diag}(+1,-1,-1,-1)$. The field is regarded as an operator-valued distribution $f\mapsto\phi(f)$ on the space $\cS(\IM)$ of Schwartz functions $f$, where the operators are all defined on a common dense and stable domain $\mathcal{D}$ of the underlying Hilbert space $\mathcal{H}$. The algebra of local observables is the $^*$-algebra $\cA(\IM)$, generated by multiples of $\eins$ and finite sums as well as products of the field operators. This algebra is stable under the action of the proper, orthochronous Poincar\'e group $\mathcal{P}_+^{\uparrow}$, implemented on the field operators by
\ben
 \t_{(\L,a)}(\phi(f))=\phi(f_{(\L,a)}),
\een
where $f_{(\L,a)}(x)=f(\L^{-1} (x-a))$, and stable under the action of the gauge group $\IZ_2$, acting as $\g(\phi(f))=-\phi(f)$. Furthermore, we assume that 
\begin{itemize}
 \item[i)] $f \mapsto \phi(f)$ is linear.
 \item[ii)] $\phi(f)^*=\phi(\bar{f})$ for all $f\in\cS(\IM)$.
 \item[iii)] Klein Gordon equation: $\phi((\square+m^2)f)=0$ for all $f\in\cS(\IM)$, where $\square$ denotes the d'Alembert operator and $m\geq 0$ is the mass parameter.
 \item[iv)] Canonical Commutation Relations (CCR): $\left[ \phi(f),\phi(g) \right]=i E(f,g)\eins$ for all $f,g\in\cS(\IM)$, where $E$ denotes the causal propagator, which is defined as the difference of the advanced minus the retarded fundamental solution of the Klein-Gordon equation. Einstein causality is expressed by $E(f,g)=0$ if $f$ and $g$ have mutually spacelike separated supports.
\end{itemize}

A \emph{state on} $\mathcal{A}(\IM)$ is a continuous normalized positive linear functional $\o:\cA(\IM)\rightarrow \IC$. The $n$-point ``functions`'' of a state are distributions in $\cS^\prime(\IM^n)$, formally given by
\ben
 \o_{n} (x_1,\ldots,x_n):=\o(\phi(x_1)\cdots\phi(x_n)),\quad n\in\IN.
\een

Mostly, we will focus on quasifree states which are determined by their two-point functions $\o_2$ through
\ben
\o\left(e^{it\phi(f)}\right)=e^{-\frac{1}{2}\o_2(f,f)\cdot t^2},
\een
where the equation is to interpreted as equating terms of equal order in $t$. Furthermore, we assume that the states are gauge invariant, which means $\o\circ\g=\o$.

In the following we will only consider (quasifree) states fulfilling the \emph{Hadamard condition}, characterized by the following restriction on $\WF(\o_2)$, the wave front set of their two-point functions,
\ben
 \WF(\o_2)=\{(x,x^\prime,k,-k)\in T^\ast\IM^2:x\sim_k x^\prime,k_0>0\},
\label{eq:WFmink}
\een
or even {analytic Hadamard states}, characterized by a restriction on the analytic wave-front set of their two-point functions:
\ben
 \WF_A(\o_2)=\{(x,x^\prime,k,-k)\in T^\ast\IM^2:x\sim_k x^\prime,k_0>0\}.
\label{eq:WFAmink}
\een
For a discussion of the properties and a definition, the reader is referred to \cite{SV00, SVW02} and references therein. For a motivation why one would prefer to consider Hadamard states, see e.g. \cite{BDH13}.

\begin{defn}
 Let $\cA$ be a $^*$-algebra, $\a_t$ a one-parameter group of automorphisms on $\cA$,  $\o$ a state on $\cA$ and $\b>0$. Define the open strip $S_\b$ by $S_\b:=\{z\in\IC: 0<\Im z<\b\}$ and denote by $\bar{S}_\b$ the closed strip. Then $\o$ is called a \emph{KMS state at value} $\b$ \emph{with respect to} $\a_t$ (or $(\b,\a_t)$-{KMS state}, for short), iff for any $A,B\in\cA$ there exists a function $F_{A,B}$, which is defined and holomorphic on $S_\b$, and continuous on $\bar{S}_\b$, with boundary values
\begin{align}
 F_{A,B}(t)&=\o\left(A\a_t(B)\right),\label{eq:KMS1}\\
 F_{A,B}(t+i\b) &= \o \left(\a_t(B)A\right)\label{eq:KMS2},
\end{align}
for all $t\in\IR$.
\end{defn}
A Lorentz frame is fixed by the choice of a future-directed timelike unit vector $e$; this means $e\in V_+$, where $V_+$ denotes the open forward lightcone, and $e^2\equiv e^\mu e_\mu=1$. The set of those vectors will be denoted by $V_+^1$ in the following. In the present model the one-parameter group of time evolution on $\cA(\IM)$ with respect to the Lorentz frame fixed by some $e\in V_+^1$ is given by
\ben
\a^{(e)}_t=\t_{(1,te)},\quad t\in\IR.
\een

A KMS state $\o_\b$ with respect to $\a_t^{(e_\b)}$ is regarded as a thermal equilibrium state at inverse temperature $\b$ with respect to the rest system (or Lorentz frame) specified by some $e_\b\in V_+^1$.  Therefore thermal equilibrium states in relativistic QFT are indicated by both inverse temperature $\b$ and time direction $e_\b$ of the rest system. It is convenient to combine the two quantities into the inverse temperature four-vector $\bb=\b e_\b \in V_+$ so that $\o_\bb$ denotes a $(\b,\a_t^{(e_\b)})$-KMS state on $\cA(\IM)$. We therefore call call $\o_\bb$ simply a $\bb$-\emph{KMS state}.
To rule out possible phase transitions, we assume that for any given $\bb$ there is a unique gauge-invariant $\bb$-KMS state $\o_\bb$ on $\cA(\IM)$. This assumption also implies that $\o_\bb$ is invariant under spacetime translations. Furthermore we point out that $\bb$-KMS states are quasifree states and fulfill the analytic microlocal spectrum condition \cite{SVW02}, in particular they are analytic Hadamard states. 

It has been shown in \cite{BB94} that the correlation functions $$F_{A,B}(x)=\o_\bb(A\t_{(1,x)}(B)),\ x\in\IR^4$$ of $\bb$-KMS states $\o_\bb$ on $\cA(\IM)$ have in fact stronger analyticity properties than those implied by the KMS condition. These analyticity properties can be seen as a remnant of the relativistic spectrum condition in the case of a thermal equilibrium state. 
\begin{defn}
 A state  $\o_\bb$ on $\cA(\IM)$ satisfies the \emph{relativistic KMS condition} at inverse temperature $\beta>0$ iff there exists some $e_\b\in V_+^1$, such that for any $A,B\in\cA(\IM)$ there exists a function $F_{A,B}$ which is defined and holomorphic in the tube $\cT_{\beta e_\b}=\{z\in\IC^4:\Im z\in V_+\cap(\b e_\b+V_-)\}$, where $V_-=-V_+$, and continuous at the boundary sets $\Im z=0$ and $\Im z=\beta e_\b$ with
\begin{align}
 F_{A,B}(x)=&\o_\bb(A\t_{(1,x)}(B)),\\
F_{A,B}(x+i\b e_\b)=&\o_\bb(\t_{(1,x)}(B)A),\quad x\in\IR^4.
\end{align}
\end{defn}

\section{The LTE condition of Buchholz, Ojima and Roos}
The first key step in the analysis of Buchholz, Ojima and Roos in \cite{BOR02} is the construction of spaces $\cQ_q$ of idealized observables (density-like quantities) located at $q\in\IM$. Those observables are well-defined as quadratic forms and their expectation values can be calculated in all states with an appropriate high-energy behaviour. From the spaces $\cQ_q$ one then selects subspaces $\cS_q\subset\cQ_q$ of local thermal observables $s(q)$. The thermal interpretation of these observables is justified by evaluating them in thermal reference states. The set of these reference states is denoted by $\cC_B$ and consists of mixtures of KMS states $\o_\bb$, with $\bb$ contained in some compact subset $B\subset V_+$. A generic state $\o_B\in\cC_B$ is represented in the form
\ben
\o_B(A)=\int_B \diff \mu(\bb) \o_\bb(A), \quad A\in\cA(\IM),
\label{eq:mixed}
\een
where $\mu$ is a positive normalized measure on $V_+$, with support contained in $B$.

The connection between the local thermal observables from the spaces $\cS_q$ and the macroscopic thermal properties of a reference state is provided as follows: As discussed explicitly in \cite{Buc03}, the local observables $s(q)$ yield the same information on the thermal properties of the reference states as certain macroscopic observables $\fS$, namely for certain sequences $f_n\in\cD(\IR^4)$ with $f_n\nearrow 1_{\IR^4}$ the limit
\ben
\fS=\lim\limits_{n\to\infty} s(f_n)
\een
exists in all thermal reference states and defines a macroscopic (central) observable, i.e. $\fS$ is commuting with any element $A\in\cA$ as well as with the spacetime translations $\t_{(1,a)},\ a\in\IR^4$. One assumes that all macroscopic intensive thermal parameters of a $\bb$-KMS state are given by maps $\bb\mapsto\fS(\bb)$ which are called \textit{thermal functions}. For any $s(q)\in \cS_q$ we can define such functions by
\ben
\bb\mapsto\fS(\bb):=\o_\bb(s(q)),
\een
which are Lorentz tensors with the tensorial character depending on $s(q)$. Furthermore, as a consequence of spacetime translation invariance of the states $\o_\bb$, they do not depend on the specific choice of the point $q\in\IM$. The thermal functions yield the central decomposition of the macroscopic observables $\fS$ \cite{Buc03}. Thus, we can identify $\fS$ with the respective thermal function $\fS(\bb)$ and the states $\o_B$ can be lifted to the space of macroscopic observables via
\ben
\o_B(\fS)(q):=\o_B(s(q)),\quad s(q)\in S_q.
\een  

In the present model the spaces of thermal observables are defined as the spaces $\cS_q^n$, spanned by the so-called \emph{balanced derivatives of the Wick square up to order} $n$. Those are defined as
\ben
\eth_{\mu_1\ldots\mu_n}:\phi^2:(q):=\lim\limits_{\xi\to 0}\del_{\xi^{\mu1}}\ldots\del_{\xi^{\mu_n}}\left[\phi(q+\xi)\phi(q-\xi)-\o_\text{vac}(\phi(q+\xi)\phi(q-\xi))\cdot\eins\right],
\label{eq:bderiv}
\een
where $\o_\text{vac}$ is the unique vacuum state on $\cA(\IM)$ and the limit is taken along spacelike directions $\xi$. Of particular interest is the space $\cS_q^2$ which contains (besides the unit $\eins$) two thermal observables which play a prominent role. The first one is $:\phi:^2(q)$, the \textit{Wick square} of $\phi$ at the point $q\in\IM$, which is usually regarded as corresponding to a point-like ''thermometer observable`` $\Th(q)$. This is due to the fact that its evaluation in a $\bb$-KMS state yields for the Klein-Gordon field with $m=0$:\footnote{In the massive case the expression $\o_\bb(:\phi^2:(q))$ yields a slightly more complicated but still monotonously decreasing function of $\b$.}
\ben
\Th(\bb):=\o_{\bb}(:\phi^2:(q))=\frac{1}{12\b^2}=\frac{k_B^2}{12} T^2.
\een
The other thermal observable contained in $\cS_q^2$ is $\eth_{\mu\nu}:\phi^2:(q)$, the second balanced derivative of $:\phi^2:(q)$. It is of special interest since its expectation values in a $\bb$-KMS state $\o_\bb$ are (up to a constant) equal to the expectation values of the thermal stress-energy tensor \cite{BOR02}:
\ben
E_{\mu\nu}(\bb):=-\frac{1}{4}\o_\bb(\eth_{\mu\nu}:\phi^2:(q))=\frac{\pi^2}{90}\left(4\bb_\mu\bb_\nu-\bb^2\eta_{\mu\nu}\right)(\bb^2)^{-3}.
\een

For the Klein-Gordon field with $m=0$ an easy computation yields \cite{BOR02}:
\ben
\fS^{(n)}(\bb):=\o_\bb(\eth_{\mu1\cdots\mu_n}:\phi^2:(q))=c_n\del_{\mu1}^\bb\ldots\del_{\mu_n}^\bb\left(\bb^2\right)^{-1}.
\een
This makes clear that the thermal functions $\fS^{(n)}(\bb)$ can be constructed completely out of $\bb$.\footnote{This is also true in the massive case. Here, the thermal functions are given by a more involved expression which is analytic in $\bb$ \cite{Hue05}.} Thus they can be viewed as thermal functions corresponding to the micro-observables $s(q)$. Furthermore, due to the invariance of $\o_\bb$ under spacetime translations, they are independent of $q$. Note, that for odd $n$ the thermal functions are equal to $0$. 

The definition of local thermal equilibrium in the sense of \cite{BOR02, Buc03} can now be stated for the quantized Klein-Gordon field\footnote{In \cite{BOR02} a definition has been given which is valid for more general quantum fields $\phi$, also including interacting ones.} as follows:

\begin{defn}
 Let $\cO\subset \IM$ and $\o$ a Hadamard state on $\cA(\IM)$. 
\begin{itemize}
\item[1.)] We say that $\o$ is a \emph{local thermal equilibrium  state of order $N$ in $\cO$ with sharp inverse temperature vector field} $\bb(\cO)$, or $[\bb(\cO),N]$-\emph{LTE state} for short, iff there exists a continuous (resp. smooth, if $\cO$ is open) map $\bb:\cO\to V_+$ for any $q\in\cO$ it holds
\ben
\o(s(q))=\o_{\bb(q)}(s(q))\quad \forall s(q)\in \cS_q^n,\ n\leq N,
\label{eq:sLTE}
\een
where $\o_{\bb(q)}$ is the unique extremal $\bb(q)$-KMS state on $\cA(\IM)$

\item[2.)] We say that $\o$ is a \emph{local thermal equilibrium state of order $N$ in $\cO$ with mixed temperature distribution $\mu$}, or $[\mu,\cO,N]$-\emph{LTE state} for short, iff there exists a function $\mu:q\mapsto \mu_q,q\in\cO$, where each $\mu_q$ is a probability measure with support in some compact $B(q)\subset V_+$, and for any $q\in\cO$ 
 \ben
 \o(s(q))=\o_{B(q)}(s(q)),\quad \forall s(q)\in S_q^n,\ n\leq N,
\label{eq:mLTE}
 \een
where the states $\o_{B(q)}, q\in\cO$ are defined by
\ben
\o_{B(q)}(A)=\int_{B(q)}d\mu_q(\bb)\o_\bb(A), \quad A\in\cA(\IM).
\een 
We say that $\o$ is a $[\mu,\cO]$-LTE state iff (\ref{eq:mLTE}) holds for all $n\in\IN$.
\end{itemize}
\label{defn:LTE}
\end{defn}

It is obvious from this definition that any $\bb$-KMS state $\o_\bb$ is a $\bb(\IM)$-LTE state with constant inverse temperature vector field given by $\bb(q)\equiv\bb$. Although this should be the case for consistency reasons the noteworthy feature of the above definition lies in the possibility of a \textit{varying} inverse temperature vector field $\bb$, so an LTE state can have varying inverse temperature $\b_q$ (resp. inverse temperature distribution $\mu_q$) as well as varying rest frame at each $q\in\cO$. 

It is known from special relativistic thermodynamics that all relevant macroscopic thermal parameters, in particular the entropy current density, for a (local) equilibrium state can be constructed once the components of $E_{\mu\nu}$ are known \cite[Chapter 4]{Dix78}. This means that in order to gain knowledge about the coarse macroscopic properties of (local) equilibrium states it is sufficient to analyze them by means of the subset $S_q^2$ of all thermal observables.  For increasing $n$ the spaces $\cS_q^n$ contain more and more elements, i.e. the higher balanced derivatives of $:\phi^2:(q)$. Thus, the $[\bb(\cO),N]$-LTE condition introduces a hierarchy among the local equilibrium states in the following sense: If we successively increase the order $N$ in this condition we obtain an increasingly finer resolution of the thermal properties of this state. For finite $N$ we obtain a measure of the deviation of the state $\o$ from complete local thermal equilibrium (which would amount to a $\bb(\cO)$-LTE state).
  
An example of a $\bb(\cO)$-LTE state on $\cA(\IM)$ (massless case), with $\cO=V_+$, has been given in \cite{BOR02}. It is a quasifree state $\o_\text{hb}$ on $\cA(\IM)$, the so-called \emph{hot bang state} defined via
\ben
\o_2^{\text{hb}}(x,y)=\frac{1}{(2\pi)^3}\int_{\IR^4} d^4p\ \frac{\eps(p_0)\delta(p^2)}{1-e^{-\g(x+y)p}}e^{-ip(x-y)},\qquad  x+y\in V_+, 
\label{eq:hotbang}
\een
where $\g>0$ is a real parameter. One finds immediately that for all $q\in V_+$
\ben
\o_{\text{hb}}(\eth_{\mu1\ldots\mu_n}:\phi^2:(q))=\o_{\bb(q)}(\eth_{\mu1\ldots\mu_n}:\phi^2:(q)),
\een
where $\o_{\bb(q)}$ is the unique extremal $\bb(q)$-KMS state with $\bb(q)=2\g q, q\in V_+$.
Thus, the state $\o_\text{hb}$ in fact is a $\bb(V_+)$-LTE state in the sense of Definition \ref{defn:LTE}. It describes the spacetime evolution of a ``heat explosion'' with infinite temperature at the tip of the forward lightcone $V_+$ which justifies the name hot bang state. For a more thorough discussion of the properties of $\o_\text{hb}$ we refer to the article by Buchholz \cite{Buc03}. Below we will see that such a state is in fact the genereic example of an infinite-order and sharp-temperature LTE state of the massless Klein-Gordon field.

\section{A local version of the KMS condition}

The LTE condition of \cite{BOR02} is based on the heuristic assumption that one should be able to obtain information about the (macroscopic) thermal properties of near-to-equilibrium states by comparing them pointwise to thermal reference states (KMS states or mixtures of such) by means of localized thermal observables. In the present model those observables were modelled by the Wick square $:\phi^2:(q)$ and its balanced derivatives. This choice has been largely motivated by the fact that the expectation value of the Wick square in equilibrium is proportional to the square of the equilibrium temperature (i.e. temperature in the sense of the 0\textsuperscript{th} law of thermodynamics). It would clearly be desirable to give further arguments for the special choice of the thermal observables in the free field case. A physical motivation, based on the investigation of the behaviour of moving detectors modelled by quantum mechanical two-level systems (Unruh detectors), has been given in \cite{Sch07}. On the mathematical side, in view of the definition of the balanced derivatives of the Wick square, Eq. (\ref{eq:bderiv}), one should be able to encode the thermal properties of an LTE state $\o$ on $\cA(\IM)$ directly on the level of the two-point functions $\o_2$. This assumption is further strengthened by observing that the correlation functions $\o_2^\bb(q\mp\xi,q\pm\xi)$ for a $\bb$-KMS state are completely determined by the expectation values $\o_\bb(\eth_{\mu_1\cdots\mu_n}:\phi^2:(q))$ for all $n\in\IN$, as discussed in \cite{BOR02}. In the following we will discuss a recent proposal by Gransee, Pinamonti and Verch \cite{GPV15} for characterizing LTE states by properties of their two-point function $\o_2$ which are similar to the KMS condition. It will turn out that under reasonable additional analyticity requirements this characterization yields the class of $[\bb(\cO),N]$-LTE states, which were introduced in the previous section.

A first observation in \cite{GPV15} is, that for any Hadamard state $\o$ on $\cA(\IM)$ and any $q\in\IM$ the ``function'' $\fw_q$, given by
\ben
\fw_q(\xi):=\o_2\left(q-\xi,q+\xi\right),\quad \xi\in\IR^4,
\label{eq:wq}
\een
can be meaningfully defined as a distribution in $\cS'(\IR^4)$. In particular, for any timelike future-pointing unit vector $e\in V_+^1$ the ``function'' $\fu_{q,e}$, defined by 
\ben
\fu_{q,e}(t):=\o_2\left(q-t e,q+t e\right),\quad t\in\IR,
\label{eq:uq}
\een
 is well-defined as a distribution in $\cS'(\IR)$. If $\o_\bb$ is a $\bb$-KMS state on $\cA(\IM)$ it follows from the spacetime translation invariance of such states that the distribution $\fu_{\bb}$, defined by
\ben
\fu_\bb(t):=\o_2^\bb\left(q-te_\b,q+te_\b\right)
\een
 is independent of the choice of the point $q\in\IM$. The role of the parameter $t$ is enlightened by the following observation: \emph{If $\o_\bb$ is a $\bb$-KMS state on $\cA(\IM)$, then for arbitrary but fixed $q\in\IM$ there is a complex function $f_\bb$, holomorphic on the strip $S_\b$, with (distributional) boundary values}
\ben
f_\bb(t)=\o_2^\bb\left(q-t e_\b,q+t e_\b\right)\quad \textrm{and} \quad f_\bb(t+i\b)=\o_2^\bb\left(q+t e_\b,q-t e_\b\right).
\label{eq:KMSrem}
\een

This is to be seen as a remnant of the KMS condition, in which the parameter $t$ plays the same role as the parameter of the one-parametric group of time evolution on $\cA(\IM)$, and where the boundary value conditions (\ref{eq:KMS1}) and (\ref{eq:KMS2}) are replaced by the weaker property (\ref{eq:KMSrem}) above. However, the above properties are surely not sufficient to imply the $\bb$-KMS condition. The main point is that these properties are valid with respect to an a priori fixed point $q\in\IM$ and therefore do not tell us anything about spacetime translation invariance of the state $\o_\bb$. Furthermore, the knowledge of the distribution $\fu_q\in\cS'(\IR)$ (which arises as the restriction of the distribution $\fw_q\in\cS'(\IR^4)$ to the set $\{\xi\in\IR^4:\xi=t e_q,t\in\IR\}$) does not completely determine the correlation functions $\o(q\mp\xi,q\pm\xi)$  but only their restrictions to timelike arguments $\xi$. However, if one makes the additional assumption that the state $\o$ fulfills the analytic Hadamard condition \eqref{eq:WFAmink}, one observes \cite{GPV15}:

\begin{obs} An analytic Hadamard state $\o$ fulfills the $\bb(q)$-LTE condition if and only if there exists a $\bb(q)\in V_+$ such that
 \ben
\fw_q(\xi)=\fw_{\bb(q)}(\xi)=\frac{1}{(2\pi)^3}\int d^4p \frac{\eps(p_0)\d(p^2-m^2)}{1-e^{-\bb(q)p}}e^{-ip\xi},
\label{eq:LKMScomp}
\een
which is to be understood in the sense of distributions.
\end{obs}
This shows that the $\bb(q)$-LTE condition together with the analytic Hadamard condition is sufficient to determine the correlation functions $\o(q\mp\xi,q\pm\xi)$ completely. As mentioned above, the respective correlation functions for the comparison equilibrium state $\o_{\bb(q)}$ are completely fixed by the expectation values $\o_{\bb(q)}(\eth_{\mu_1\ldots\mu_n}:\phi^2:(q))$ for any $q\in\cO$. This provides an additional justification for the use of the balanced derivatives as the thermal observables in the present model. Analyzing the analyticity properties of the correlation functions $\fw_{\bb(q)}(\xi)$, the above observation is used in \cite{GPV15} to relax the KMS condition as follows:

\begin{defn}
\label{defn:LKMS}
 Let  $q\in\IM$ and $\o$ an analytic Hadamard state on $\cA(\IM)$. We say that $\o$ fulfills the \emph{local KMS condition at $q$ with respect to $\bb(q)$}, or $\bb(q)$-\emph{LKMS condition} for short, iff there exists a $\bb(q)\in V_+$ and a complex function $F_q$ with the following properties:
\begin{itemize}
\item[(i)] $F_q$ is defined and holomorphic in the (flat) tube 
\ben
\cT_q=\{z\in\IC^4:\Im z=\s e_q, 0<\s<\b_q\}.
\een
\item[(ii)] For all compact $K\subset(0,\b)$ there exist constants $C_K>0$ and $N_K\in\IN_0$ such that
\ben
\abs{F_q(\xi+i\s e_q)}\leq C_K(1+\abs{\xi+i\s e_q})^{N_K},\quad \xi\in\IR^4,\s\in K.
\label{eq:LKMS1}
\een
\item[(iii)] We have in the sense of distributions:
\begin{align}
 F_q(\xi+i\s e_q)&\xrightarrow[\s\to 0^+]{}\fw_q(\xi),
\label{eq:LKMS2}\\[1ex]
 F_q(\xi+i(\b_q-\eta)e_q))&\xrightarrow[\eta\to 0^+]{}\fw_q(-\xi),
\label{eq:LKMS3}
\end{align}
\item[(iv)] We have the following clustering property:
\ben
\fw_q(te_q)\xrightarrow[\abs{t}\to\infty]{} 0.
\label{eq:clust}
\een
\end{itemize}
 Let $\cO$ be a spacetime region. We say that $\o$ fulfills the $\bb(\cO)$-\emph{LKMS condition}, iff there exists a continuous (resp. smooth, if $\cO$ is open) map $\bb:\cO\to B\subset V_+$ such that $\o$ fulfills the $\bb(q)$-LKMS condition for all $q\in\cO$. 
\end{defn}

Any $\bb$-KMS state $\o_\bb$ is a $\bb(\cO)$-LKMS state with $\cO=\IM$, where $\b_q\equiv\b$ and $e_q\equiv e_\bb$ are constant throughout Minkowski spacetime. However, the natural question arises if there are other examples of nontrivial LKMS states. We first note that the $\bb(q)$-LKMS condition can be shown \cite{GPV15} to have an equivalent momentum-space formulation: \textit{A state $\o$ on $\cA(\IM)$ fulfills the $\bb(q)$-LKMS condition if and only if there exists a $\bb(q)=\b_q e_q\in V_+$, such that in the sense of distributions}
\ben
\wfou_q(p)=e^{\bb(q) p}\wfou_q(-p),
\label{eq:LKMSfou}
\een
\textit{and the cluster property \eqref{eq:clust} holds.}

The relation \eqref{eq:LKMSfou} can be seen as a remnant of the $\bb$-KMS condition in momentum space \cite{BB96}. With the definition \eqref{eq:hotbang} of the hot-bang state $\o_\text{hb}$ one sees that the latter is an example of a $\bb(\cO)$-LKMS state with $\cO=V_+$ and $\bb(q)=2\g q$. Thus, the local KMS condition appears as a non-trivial generalization of the KMS condition. More generally, relations \eqref{eq:clust} and \eqref{eq:LKMSfou} yield
\ben
\wfou_q(p)=\frac{1}{2\pi} \frac{\eps(p_0)\d(p^2-m^2)}{1-e^{-\bb(q)p}}=\wfou_{\bb(q)}(p)
\een

This shows that the $\bb(q)$-LKMS condition (in position or in momentum space) is sufficient to completely determine the correlation functions $\fw_q(\xi)$. In  consequence, this proves:

\begin{thm}
\label{prop:LKMS}
 Let $q\in\IM$ and $\o$ an analytic Hadamard state on $\cA(\IM)$. Then the following are equivalent:
\begin{itemize}
 \item[(i)] $\o$ is a $\bb(q)$-LTE state.
\item[(ii)] $\o$ fulfills the $\bb(q)$-LKMS condition.
\end{itemize}
\end{thm}
\paragraph{LKMS and finite-order LTE states}

For LTE states of finite order (in the sense of Definition \ref{defn:LTE}) it seems to be clear that the relation (\ref{eq:LKMScomp}) will not be valid exactly, but that a similar relation might hold. Informally, in view of the definition of the balanced derivatives (\ref{eq:bderiv}), one would expect the following to hold: \emph{The directional derivatives with respect to $\xi$ of the correlation functions $\o(q\mp\xi,q\pm\xi)$ at the point $\xi=0$ coincide with those of the respective correlation functions of a comparison equilibrium state $\o_{\bb(q)}$, up to order $N$.}

Of course, from a mathematical point of view, this statement is  meaningless, because the correlation functions are distributions in $\cS'(\IR^4)$ and it is not clear what is meant by ``the directional derivatives of $\o(q\mp\xi,q\pm\xi)$ at the point $\xi=0$''. Nevertheless, one has a mathematically well-defined version of the above informal statement \cite{GPV15}:

\begin{obs}
 Let $q\in\IM$. An analytic Hadamard state $\o$ on $\cA(\IM)$ fulfills the $[\bb(q),N]$-LTE condition if and only if there exists a $\bb(q)\in V_+$ such that
\ben
[\del_{\underline{\a}}(\fw_{\bb(q)}-\fw_q)](0)=0\quad \forall \underline{\a}\in\{\underline{\a}\in \IN_0^4:\abs{\underline{\a}}\leq N\},
\een
where $\fw_{\bb(q)}(\xi)=\o_2^{\bb(q)}(q-\xi,q+\xi)$ for the unique $\bb(q)$-KMS state $\o_{\bb(q)}$.
\end{obs}

This observation can be used to further generalize the $\bb(q)$-LKMS condition:

\begin{defn}
 Let $q\in\IM$ and $N\in\IN$. An analytic Hadamard state $\o$ on $\cA(\IM)$ is said to fulfill the $[\bb(q),N]$-\emph{LKMS condition} iff there exists a $\bb(q)\in V_+$ such that there is a complex function $F_q$ with the following properties:
\begin{itemize}
 \item[(i)] $F_q$ is defined and holomorphic in the (flat) tube 
\ben
\cT_q=\{z\in\IC^4:\Im z=\s e_q, 0<\s<\b_q\}.
\een
 \item[(ii)] For all compact $K\subset(0,\b_q)$ there exist constants $N_K\in\IN$ and $C_K>0$ such that
\ben
 \abs{F_q(\x+i\s e_q)}\leq C_K(1+\abs{\xi+i\s e_q})^{N_K},\qquad \forall\ \s\in K.
 \label{eq:LKMSN1} 
\een
 \item[(iii)] There exists a symmetric $R_q\in\cS'(\IR^4)$ with $\WF_A(R_q)=\emptyset$ and
\begin{align} 
[\del^{\underline{\a}} R_q(0)]=0 \quad&\forall \underline{a}\in\{\underline{\a}\in\IN_0^4:\abs{\underline{a}}\leq N\},\label{eq:Rq2}\\[1ex]
(\fw_q+R_q)&(te_q)\xrightarrow[\abs{t}\to\infty]{} 0 \ ,\label{eq:Rq3}
\end{align}
such that in the sense of distributions
\begin{align}
 F_q(\xi+i\s e_q)&\xrightarrow[\s\to 0^+]{}(\fw_q+R_q)(\x),
 \label{eq:LKMSN2}
\\[1ex]
 F_q(\xi+i(\b_q-\eta)e_q))&\xrightarrow[\eta\to 0^+]{}(\fw_q+R_q)(-\x).
 \label{eq:LKMSN3}
\end{align}
\item[(iv)] We have the following cluster property:
\ben
(\fw_q+R_q)(te_q)\xrightarrow[\abs{t}\to\infty]{}0.
\label{eq:clust2}
\een
\end{itemize}
\end{defn}

This definition can also be generalized to open regions $\cO$ of Minkowski spacetime. An analogous analysis as for the $\bb(q)$-LKMS condition shows that the $[\bb(q),N]$-LKMS condition is sufficient to determine the correlation functions $\o(q\mp\xi,q\pm\xi)$, similar to (\ref{eq:LKMScomp}), but only up to some real-analytic ``rest term'' $R_q:\IR^4\to\IR^4$. Without going into details, we want to state that this implies the following result:

\begin{thm}
\label{thm:LKMSN}
 Let $q\in\IM$ and $\o$ an analytic Hadamard state on $\cA(\IM)$. Then the following are equivalent:
\begin{itemize}
 \item[(i)] $\o$ is a $[\bb(q),N]$-LTE state. 
\item[(ii)] $\o$ fulfills the $[\bb(q),N]$-LKMS condition.
\item[(iii)]  There exists a $\bb(q)=\b_q e_q\in V_+$ and a symmetric $\hat{R}_q\in\cS'(\IR^4)$ with
\begin{align}
\cF[p^{\underline{\a}} \hat{R}_q(p)](0)=0 \quad&\forall \underline{\a}\in\{\underline{\a}\in\IN_0^4:\abs{\underline{\a}}\leq N\},\label{eq:Rfou1}\\[1ex]
\end{align}
 such that the cluster property \eqref{eq:clust2} holds and we have in the sense of distributions:
\ben
e^{\bb(q)p}(\wfou_q+\hat{R}_q)(-p)=(\wfou_q+\hat{R}_q)(p).
\een
\end{itemize}
\end{thm}
\noindent For a proof of Theorem \ref{thm:LKMSN} we again refer to \cite{GPV15}.

\subsubsection*{Constraints from the Klein-Gordon equation} A further interesting question is the following: Given a $[\bb(\cO),N]$-LKMS state for some (open) subset $\cO$, is the form of the map $\bb:\cO\to V_+$ completely arbitrary? Surely, this is not the case if the comparison equilibrium states $\o_{\bb(q)}$ ought to fulfill the (relativistic) KMS condition. It turns out (cf. also \cite{Buc03, Hue05}) that the equations of motion for the field $\phi$ imply dynamical constraints on the correlation functions $\fw_q(\xi)$ which give restrictions on the map $\bb$. For the case of the massless Klein-Gordon field on can prove the following:
\begin{prop}
 Let $\o\subset\IM$ and $\o$ a analytic Hadamard state on $\cA(\IM)$ which fulfills the $\bb(\cO)$-LKMS condition. Then  there exists a $\boldsymbol{b}\in\IR^4$ such that either: i) $\cO\subset\{V_+-\boldsymbol{b}\}$ (resp. $\cO\subset \{-V_+-\boldsymbol{b}\}$) and
\ben
\bb^\mu(q)=c_\o q^\mu+\boldsymbol{b}^\mu \quad \forall q\in\cO,
\een
where $c_\o>0$ (resp. $c_\o<0$) is a state-dependent constant, or ii) $\bb(q)=\text{const.}=\bb$ for all $q\in\cO$.
\end{prop}
If we exclude the somewhat unphysical case $c_\o<0$ this makes clear that the hot-bang state $\o_{\text{hb}}$, defined by \eqref{eq:hotbang}, is  the generic example of a $\bb(\cO)$-LKMS state with varying temperature. Namely, the analytic Hadamard condition on a $\bb(\cO)$-LKMS state $\o$ implies that $\fw_q=\fw_{\bb(q)}$ for all $q\in\{V_+-\boldsymbol{b}\}$ and, in consequence, that $\o$ fulfills the $\bb(\{V_+-\boldsymbol{b}\})$-LKMS condition, with 
\ben
\bb^\mu(q)=c_\o q^\mu+\boldsymbol{b}^\mu\quad \forall q\in\{V_+-b\}.
\een
The hot-bang state then corresponds to $\boldsymbol{b}=0$ and any other $\bb(\cO)$-LKMS state arises from $\o_\text{hb}$ by $\o=\o_\text{hb}\circ \t_{(1,-\boldsymbol{b})},\quad \boldsymbol{b}\in\IR^4$.

For the massive Klein-Gordon field the situation is even more restrictive: In this case it turns out that for $\cO\subset\IM$ the only states which can fulfill the $\bb(\cO)$-LKMS condition are the states for which $\bb(q)=\text{const.}$ for all $q\in\cO$. The analytic Hadamard condition on $\cO$ then implies that $\o$ has to be the unique $\bb$-KMS state $\o_\bb$ on $\cA(\IM)$. Thus, there are no nontrivial infinite-order LTE states of the massive Klein-Gordon field.
\subsubsection*{LTE states with mixed temperature} 

The above discussion implies that states of the massive Klein-Gordon field which are thermal in a subset $\cO\subset \IM$ always have to be mixed-temperature LTE states in the sense of Def. \ref{defn:LTE}, characterized at each $q\in\cO$ by some probability measure $\mu_q$. In \cite{Hue05} H\"ubener succeeded in constructing a specific example of a $[\m_q,\cO]$-LTE state. Apart from this, one has the following general existence result \cite{BOR02}:
\begin{prop}
Let $q\in\IM$. For every finite-dimensional subspace $S_q^{N}$ of all thermal observables and any compact $B_q\subset V_+$ there exists a probability measure $\mu_q$, with support contained in $B_q$, and states $\o$ on $\cA(\IM)$ which are $[\mu_q,N]$-thermal.
\end{prop}

This result has been generalized by Solveen \cite{Sol10} as follows:
\begin{prop}
Let $\cO$ be a compact region of Minkowski spacetime. For every finite-dimensional subspace $S_q^{N}$ of all thermal observables there exists a map $\mu:q\mapsto\mu_q,q\in\cO$, where $\mu_q$ is a probability measure compactly supported in $V_+$, and states $\o$ on $\cA(\IM)$ which are $[\mu(\cO),N]$-thermal.
\end{prop}
In view of these existence results,  it seems to be desirable to give an intrinsic characterization of such states similar to the LKMS condition. Similar to the case of sharp-temperature LTE states one observes the following \cite{GPV15}:

\begin{obs}
 Let $q\in\IM$. An analytic Hadamard state $\o$ on $\cA(\IM)$ fulfills the $[\m,\{q\},N]$-LTE condition if and only if there exists a probability measure $\mu_q$ with support in some compact $B(q)\subset V_+$, such that
\ben
[\del_{\underline{\a}}(\fw_{B(q)}-\fw_q)](0)=0 \quad \forall \underline{\a}\in\{\underline{\a}\in\IN_0^4:\abs{\underline{\a}}\leq N\},
\een
where $\fw_{B(q)}(\xi)=\int_{B(q)}d\mu_q(\bb)\fw_\b(\xi)$.
\end{obs} 
\noindent Unfortunately, one immediately obtains that, although the distribution $\fw_{B(q)}$ can be extended to a holomorphic function on a subset of $\IC^4$, it does not have periodicity properties in the imaginary space-time variable, since the state $\o_{B(q)}$ does not fulfill the KMS condition with respect to some $\bb\in V_+$. However, there might be the possibility to characterize such states by remnants of the so-called \emph{auto-correlation inequalities}, which yield another (equivalent) characterization of equilibrium states in algebraic quantum statistical mechanics (see e.g. \cite[Thm. 5.3.15 and Thm. 5.3.17]{BR97}). This problem is currently under investigation.

\section*{Summary and Outlook}
In this article we reviewed some aspects of local thermal equilibrium states in relativistic quantum field theory. The necessity to introduce such states arises since one would like to describe the macroscopic properties of states in quantum field theory which are not global equilibrium (KMS) states, but locally still possess well-defined thermal parameters, like temperature and thermal stress-energy. For the characterization of LTE states of the quantized Klein-Gordon field on Minkowski spacetime one has in principle two options. One could describe these states in  operational way, as it has been done in \cite{BOR02}, which results in an (extrinsic) LTE condition. On the other hand, one could aim at a more intrinsic characterization, based on properties of correlation functions, in the spirit of the KMS condition. Such a generalized KMS condition, called local KMS condition, has been introduced in \cite{GPV15}, and it turns out that, under additional (physically motivated) analyticity assumptions on the two-point function, both approaches yield the same class of non-equilibrium states of the quantized Klein-Gordon field on Minkowski spacetime. 

Finally, we want to mention that the concept of LTE states has also been generalized to include quantum fields on a generic curved spacetime \cite{BS07, SV08, Sol12} and some results concerning the thermal behaviour of quantum fields in cosmological spacetimes of Friedmann-Robertson-Walker type \cite{Wal84} have been established \cite{Gra10, Kno10, Sch10}. For an overview and a more in-depth discussion of these results and other results concerning LTE states in quantum field theory, we refer the interested reader to the exhaustive review article by Verch \cite{Ver12} and the references therein. It clearly is a challenging task to try to generalize the results concering the LKMS condition also to situations in which gravity is present, i.e. in which space-time is curved.

\subsubsection*{Acknowledgements} The author wants to thank the organizers of the conference ``Quantum Mathematical Physics'' for their invitation and kind hospitality during his stay in Regensburg. Parts of the work on the LKMS condition have been carried out during a stay at the ESI in Vienna for the conference ``Algebraic Quantum Field Theory - Its status and its future'' in May 2014. Financial support from the International Max Planck Research School ``Mathematics in the Sciences'' is gratefully acknowledged.

\bibliographystyle{amsplain}

\providecommand{\bysame}{\leavevmode\hbox to3em{\hrulefill}\thinspace}
\providecommand{\MR}{\relax\ifhmode\unskip\space\fi MR }
\providecommand{\MRhref}[2]{%
  \href{http://www.ams.org/mathscinet-getitem?mr=#1}{#2}
}
\providecommand{\href}[2]{#2}

\end{document}